# Assessing Bilateral Neurovascular Bundles Function with Pulsed Wave Doppler Ultrasound: Implications for Reducing Erectile Dysfunction Following Prostate Radiotherapy

Jing Wang[1], Xiaofeng Yang[2], Boran Zhou[3], James Sohn[4], Richard Qiu[2], Pretesh Patel[2], Ashesh B. Jani[2], and Tian Liu[1*]

[1]Department of Radiation Oncology, Icahn School of Medicine at Mount Sinai, New York, NY
[2]Department of Radiation Oncology and Winship Cancer Institute, Emory University, Atlanta, GA
[3]Department of Radiation Oncology, Medstar Georgetown University Hospital, Washington, D.C.
[4]Department of Radiation and Cellular Oncology, The University of Chicago, Chicago, IL

**Abstract:** This study aims to evaluate the functional status of bilateral neurovascular bundles (NVBs) using pulsed wave Doppler ultrasound in patients undergoing prostate radiotherapy (RT). Sixty-two patients (mean age: 66.1 ± 7.2 years) underwent transrectal ultrasound scan using a conventional ultrasound scanner, a 7.5 MHz bi-plane probe and a mechanical stepper. The ultrasound protocol comprised 3 steps: 1) 3D B-mode scans of the entire prostate, 2) localization of NVBs using color flow Doppler imaging, and 3) measurement of NVB function using pulsed wave Doppler. Five pulsed Doppler waveform features were extracted: peak systolic velocity (PSV), end-diastolic velocity (EDV), mean velocity (Vm), resistive index (RI), and pulsatile index (PI). A phantom experiment was conducted to validate the accuracy of the blood flow measurement. For 62 patients, the Doppler pulsed waveform parameters ranged as follows: PSV = 4.81 – 34.94 cm/s, EDV = 0 – 12.15 cm/s, Vm = 0.86 – 17.66 cm/s, RI = 0.27 – 1.0, and PI = 0.39 – 11.69. Blood flow discrepancy analysis of PSV revealed varying degrees of difference between left and right NVBs: 43 of 62 patients exhibited less than 50% difference, 12 patients displayed between 50 to 100% difference, and 7 patients showed over 100% difference. In a subset of patients younger than 65 years (n=13) with available EPIC sexual function scores, a correlation between decreased blood flow and sexual dysfunction was observed (Spearman correlation coefficient of -0.81, $p < 0.01$). In summary, this study presents a Doppler evaluation of NVBs in patients undergoing prostate RT. It highlights substantial differences in Doppler ultrasound waveform features between bilateral NVBs. The proposed ultrasound method may prove valuable as clinicians strive to deliver NVB-sparing RT to preserve sexual function effectively and enhance patient overall well-being.

# INTRODUCTION

Impairment of sexual potency is a common complication of radiotherapy treatment (RT) for prostate cancer, affecting a significant proportion of patients (1, 2). Studies estimate that approximately 50% of men will experience post-radiation erectile dysfunction (ED) (3). Despite its prevalence, the precise mechanism behind radiation-induced ED remains elusive. However, research suggests that ED following RT is correlated with radiation dose to various structures, including the bulb of the penis, the crura, and notably, the neurovascular bundle (NVB) (4-7). The bilateral NVBs, situated along the postero-lateral aspect of the prostate, have emerged as the structure most strongly associated with RT-related ED (8-10). While NVB-sparing RT has shown promise in terms of contour reproducibility (11) and treatment planning feasibility (12), it is not currently a standard avoidance structure in prostate RT protocols. This underscores the pressing need for imaging technologies that can facilitate its incorporation into treatment planning.

In the field of prostate imaging, standard modalities such as MRI, CT, and ultrasound are well-established and widely employed (13). However, there has been a noticeable lack of imaging studies focusing specifically on the NVB or radiation-induced NVB injury (14). While some previous research has explored ultrasound imaging, particularly utilizing Doppler ultrasound velocity waveform indices for quantitative assessments of arterial blood flow (15-20), further investigations in the context of prostate RT are lacking. Doppler ultrasound with clinically accepted parameters in terms of peak systolic velocity (PSV), end-diastolic velocity (EDV), and resistive index (RI), provides data of vascular state (21, 22). For example, a recent study showed that the EDV and RI measured at bilateral NVB vessels through Doppler transrectal ultrasound were associated with prostate cancer (23). However, this work has not been followed with further exploratory technical or clinical studies in the area of prostate RT. To date, a few of NVB blood flow studies were conducted in the prostate biopsy or surgery settings(23, 24).

Addressing these limitations, we have conducted a Doppler ultrasound imaging study to evaluate NVB blood flow in patients receiving external-beam and/or high-dose-rate RT for prostate cancer. We first conducted a flow phantom study to validate the accuracy of the Doppler parameter measurements, followed by a prospective patient study. Through this endeavor, we aim to contribute to the refinement of imaging methodologies for NVB assessment. Pulsed wave Doppler offers physicians valuable insights into the presence or absence of flow in the NVBs, which can be instrumental in enhancing unilateral or bilateral NVB-sparing treatment strategies for prostate cancer patients.

## METHOD AND MATERIALS

**Phantom experiment to assess the accuracy of the Doppler velocity measurement**

The flow phantom experiments were performed to assess the accuracy of Doppler velocity measurements using precision flow rates and a proprietary blood-mimicking fluid. A Doppler flow phantom (Model 403, Sun Nuclear Corporation, sound speed = 1550 ± 10 m/s) was utilized in this study (Fig. 1). The vessel diameter was 5 mm, and flow speed was measured at a vessel depth of 3.9 cm. The Doppler angle was set at 51º. Flow speed in the vessel was measured using an ultrasound system (Arietta 70, Hitachi Aloka Medical America, Inc.) and a 7.5 MHz bi-plane probe (EUP-U533C). The measurements were conducted in both continuous and pulsatile flow modes, with three different flow rates (2, 5, 8 mL/s) generated for the ultrasound dataset. This phantom experiment served as a calibration process to ensure the accuracy of our Doppler ultrasound systems for precise flow measurements, optimizing output and providing meaningful results for patients undergoing NVB blood flow examinations.

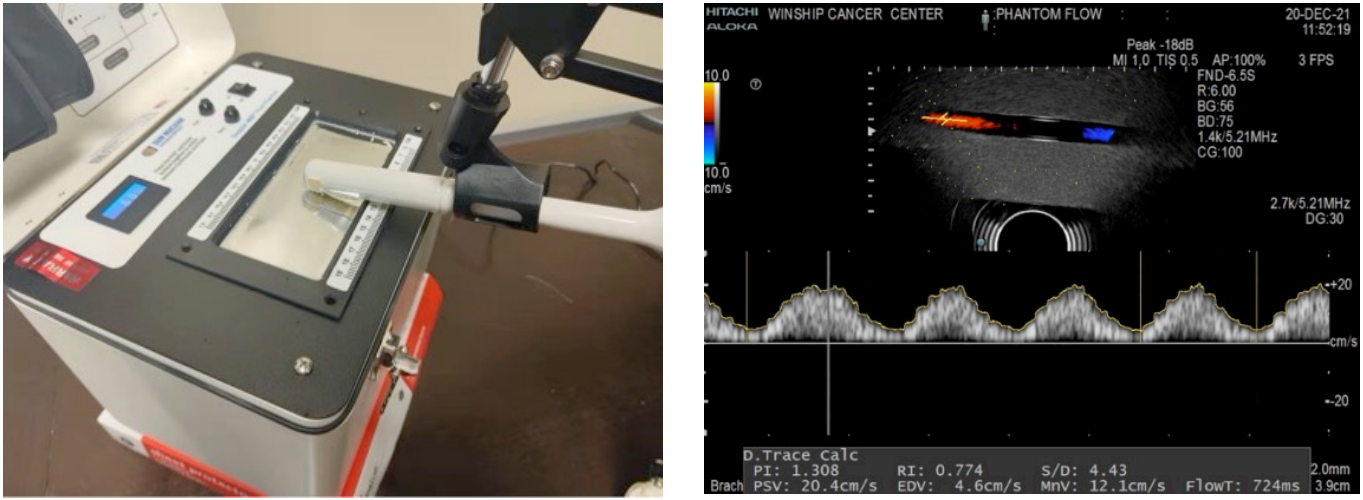

**Figure 1.** (a) Setup for Doppler flow phantom measurements, and (b) Pulsed wave Doppler measurement of the vessel within the flow phantom.

**Patient Study – Doppler Ultrasound of the NVBs**

In an IRB-approved clinical study, patients diagnosed with prostate cancer and treated with external-beam RT and/or high-dose-rate brachytherapy were recruited. Each patient received transrectal ultrasound imaging of the prostate using a HI VISION Avius ultrasound machine with a bi-plane probe for transverse and sagittal scans. During data acquisition, a mechanical support arm (Bard Medical, Inc, Georgia, USA) was used to hold and manipulate the transrectal probe.

The ultrasound protocol comprised three steps. Step 1, as depicted in Figure 2, involved scanning the patients in the lithotomy position, with 3D B-mode transverse scans captured from the apex to the base with a 2 mm step size to cover the entire prostate gland. All ultrasound B-mode images were acquired with settings: 7.5-MHz center frequency, 3.3-cm focal length, 5-cm depth, 65 brightness gain, and 18 frames per second. Before measuring Doppler parameters, patients received the 3D B-mode images, following the same procedures described previously (25). The voxel size of the 3D ultrasound dataset was 0.12×0.12×1.00 mm³. During the ultrasound scan, care was taken to avoid excess probe pressure on the rectal wall. Additionally, each patient was asked to empty their bladder to prevent compression on the prostate and NVBs. Step 2 involved localization of NVBs using color flow Doppler imaging by a radiation oncologist. Step 3 consisted of measurement of NVB function using pulsed wave Doppler.

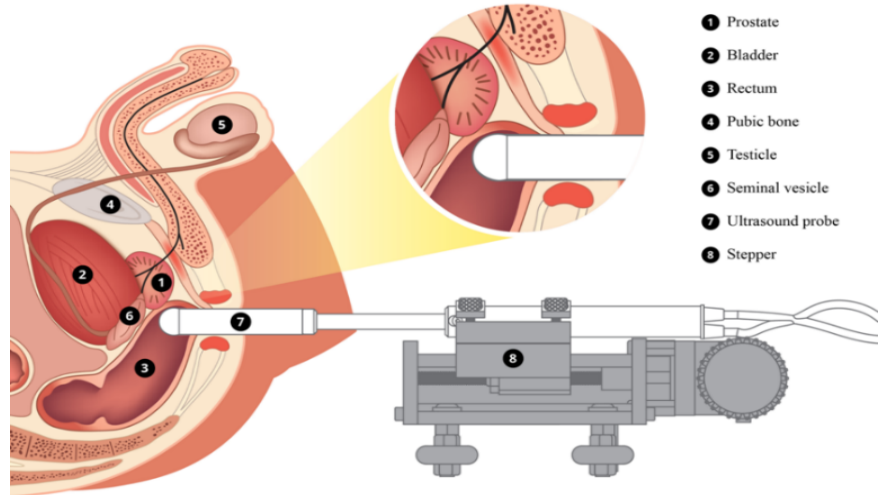

**Figure 2.** Diagram illustrating 3D B-mode and Doppler ultrasound scans. The patient is scanned in a lithotomy position using a probe. The probe angle is controlled using the SurePoint stepper set at a fixed angle.

Subsequently, Doppler spectrum waveform parameters were measured to determine the NVB blood flow. The prostatic vasculatures at bilateral NVB sites were analyzed in the largest transverse section of the prostate in accordance with reports by Leventis *et. al.* (26) and Neumaier *et. al.* (27). Five pulsed Doppler waveform features were extracted: peak systolic velocity (PSV), end-diastolic velocity (EDV), mean velocity (Vm), resistive index (RI), and pulsatile index (PI).

**Pulse waveform feature extraction**

Automatic Doppler spectral envelop detection was performed using an in-house program using Python (version 3.7.15). The patients' Doppler ultrasound images were automatically cropped to only focus on the Doppler signal. The signal was first renormalized to brighten the pixels and obtain a better view of doppler waveform. Then an intensity thresholding (40% of the maximum brightness) was adopted to identify the pixel boundary of doppler spectra, and a Gaussian filter was used for spectral smoothing of the doppler waveform. Differences in the Doppler signal have been quantified by measuring discrepancies in extracted velocity parameters, for example, PSV, EDV, and Vm. Waveform feature extraction is the determination of a feature or a feature vector from a pattern vector (28), which captures the blood flow characteristics in the time and spectral domain while independent of the Doppler angle (29).

**Statistical analysis**

Descriptive statistics will be first employed to summarize the mean and standard deviation of the Doppler parameters between right and left NVBs at each US image, respectively. Paired t-test will be further conducted to investigate the unilateral difference in the Doppler parameters between left and right NVB, respectively.

## RESULTS

**1. Flow Phantom Results**

Both continuous and pulsatile modes of the flow phantom was used to collect US doppler data. According to the volume flow rates and vessel diameter, we calculated the PSVs in phantom. From the acquired doppler images, we obtained PSVs from the pulsed waves. Table 1 shows the results of computed and doppler PSVs for 2, 5, and 8 mL/s flow rates, at both continuous and pulsatile modes.

**Table 1.** Doppler Velocity PSV Measurement Accuracy Compared with Computation from Phantom Setting.

| Phantom Mode | Volume flow rate in phantom (mL/sec) | Computed PSV in phantom (cm/s) | Doppler PSV (cm/s) | Deviation between computed and measured PSVs |
|---|---|---|---|---|
| Continuous | 2 | 20.4 | 21.9 | 6.8 % |
| | 5 | 50.9 | 51.4 | 0.9 % |
| | 8 | 81.4 | 78 | 4.2 % |
| Pulsatile | 2 | 20.4 | 20.9 | 2.3 % |
| | 5 | 50.9 | 51.4 | 0.9 % |
| | 8 | 81.4 | 78 | 4.2 % |

## 2. Patient Results

In this report, imaging data from 62 patients (66.1 ± 7.2 years) were analyzed. Patient characteristics are illustrated in Table 2. For 62 patients, the Doppler pulsed waveform parameters ranged as follows: PSV = 4.81 – 34.94 cm/s, EDV = 0 – 12.15 cm/s, Vm = 0.86 – 17.66 cm/s, RI = 0.27 – 1.0, and PI = 0.39 – 11.69.

**Table 2.** Patient characteristics.

| Characteristics | Value |
|---|---|
| *Age (y)* | |
| Mean ±std | 66.1 ± 7.2 |
| *Race* | |
| African-American | 34 |
| White | 28 |
| *Received External Beam Radiation Therapy* | |
| Yes | 33 |
| No | 29 |
| *Patients' Gleason Grade (Prostate Cancer)* | |
| 6 | 4 |
| 7 | 41 |
| 8 | 8 |
| 9 | 9 |

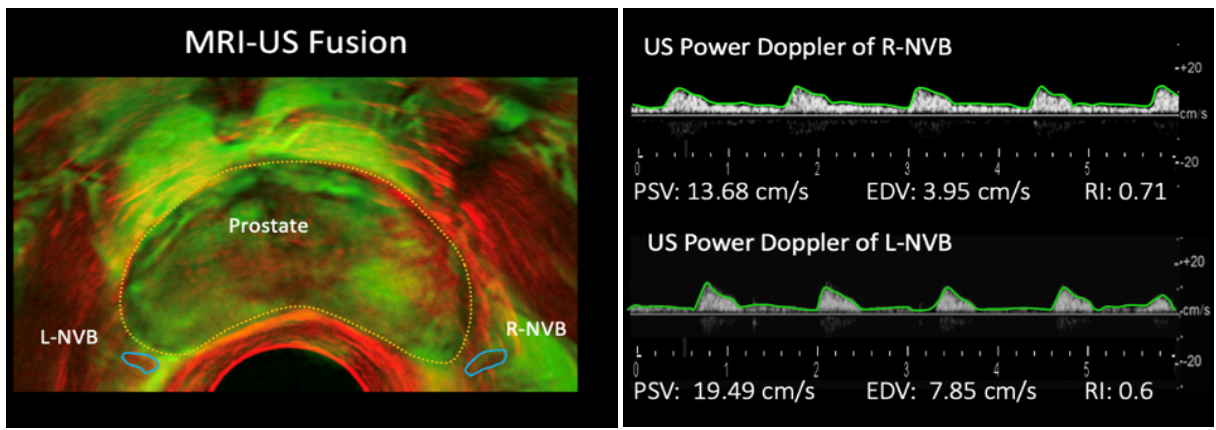

**Figure 3.** (a) MRI-US fusion of the prostate indicating the locations of the bilateral NVBs, and (b) Doppler ultrasound measurements of the left and right NVBs.

Bilateral NVBs were localized via color flow US imaging and confirmed with MRI-US fusion (Figure 3). The discrepancy in blood flow between left and right NVBs was assessed as a percentage difference ([max -

min]/min x 100. Large variations in the unilateral discrepancy of doppler US waveform features between left and right NVB were observed (Table 3).

Table 3. Discrepancy in the pulsed wave features between left and right NVBs (n=62).

|  | PSV | EDV* | Vm | RI | PI |
|---|---|---|---|---|---|
| Unilateral discrepancy [%] | 47 ± 54 | 737 ± 978 | 84 ± 105 | 26 ± 46 | 98 ± 124 |

* The percentage discrepancy calculations for EDV were based on only 17 cases after excluding those with an EDV value of 0.

### 3. Doppler parameters correlating with sexual function

Among the 62 enrolled patients, 31 of them (mean age 66 ± 8, range 51 – 83 years) self-reported their sexual function using the Expanded Prostate Cancer Index Composite (EPIC) questionnaire, from which erectile scores were obtained and compared with the Doppler parameters. The EPIC erectile scores range from 0 to 12, with lower scores indicating better erectile function. We found that in the younger group (age ≤ 65 years, 13 cases), the mean PSV value of the left and right sides was negatively related to the erectile score, with a Spearman correlation coefficient of -0.81 ($p < 0.01$). However, in the older group (age > 65 years, 18 cases), no Doppler parameter was identified to be significantly correlated with the erectile score ($p < 0.05$). Figure 4 illustrates the relationship between the PSV mean value and the erectile score for the younger group and the older group.

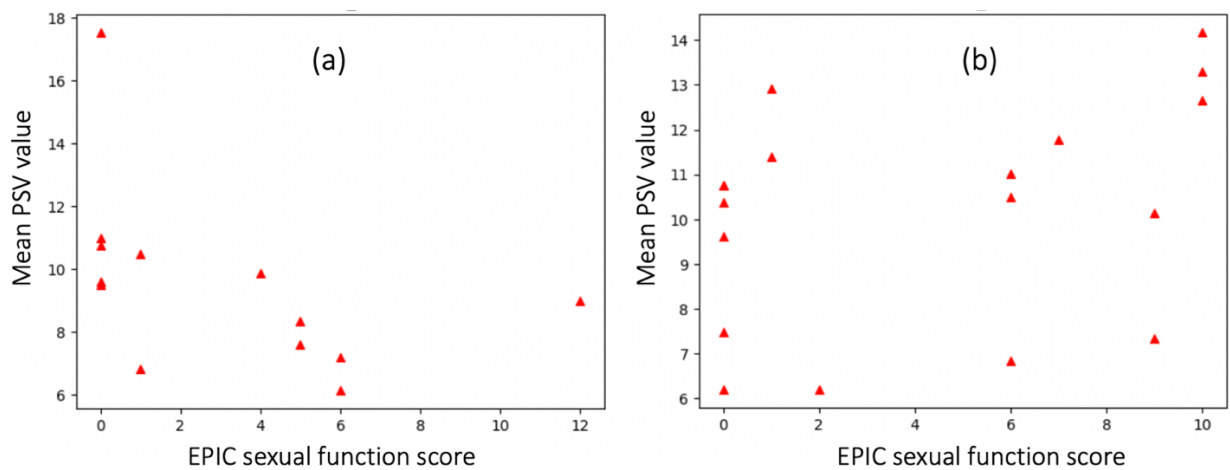

**Figure 4.** Mean PSV value of left and right NVBs versus the erectile score of patients in two groups: (a) patients aged <= 65 years and (b) patient aged > 65 years.

# DISCUSSIONS

1. **Significance of Findings**:

The results of this study provide valuable insights into the functional assessment of NVBs using Doppler ultrasound in the context of prostate RT. By examining Doppler waveform features such as PSV, EDV, Vm, RI, and PI, the study elucidates the variations in blood flow between the left and right NVBs. Notably, the observed discrepancies in blood flow between bilateral NVBs highlight the complexity of vascular dynamics in the prostate region, which may have implications for treatment planning and outcomes in prostate cancer patients undergoing RT.

Moreover, the correlation analysis between Doppler parameters and sexual function scores adds another dimension to the study findings. The negative correlation between PSV values and erectile scores in younger patients suggests a potential link between blood flow dynamics in NVBs and sexual function outcomes post-RT. This finding underscores the importance of considering vascular health in treatment decision-making and underscores the potential role of Doppler ultrasound as a tool for predicting sexual function outcomes in this patient population.

2. **Clinical Implications**:

The findings of this study have several clinical implications. Firstly, the observed variations in Doppler waveform features between bilateral NVBs underscore the importance of individualized treatment planning in prostate RT. Clinicians should consider the functional status of NVBs alongside anatomical considerations to optimize treatment outcomes, particularly in preserving sexual function. Additionally, the correlation between Doppler parameters and sexual function scores suggests that Doppler ultrasound may serve as a non-invasive tool for assessing vascular health and predicting post-RT erectile function outcomes (30). Integrating Doppler ultrasound into routine clinical practice could facilitate personalized treatment strategies and improve patient outcomes in prostate cancer management.

There has been a growing interest in partial or focal based therapies for prostate cancer to limit sexual and other toxicities. Real-time doppler ultrasound can be used to identify the NVB during prostate brachytherapy. Real time optimization during high-dose-rate or low-dose-rate brachytherapy planning allows an opportunity to limit dose to unilateral or bilateral NVBs and potentially decrease erectile side effects.

3. **Limitations and Future Directions:**

Despite the valuable insights provided by this study, several limitations warrant consideration. Firstly, the sample size, particularly in the subset of patients with sexual function data, is relatively small, limiting the generalizability of the findings. Future studies with larger cohorts are needed to validate the observed correlations and explore potential confounding factors. Additionally, the study primarily focused on Doppler waveform analysis, and other imaging modalities such as MRI were not utilized for comprehensive assessment. Integrating multi-modal imaging approaches could provide a more comprehensive understanding of NVB function and its implications for treatment outcomes.

Furthermore, while Doppler ultrasound shows promise as a tool for assessing NVB function, its operator-dependency and susceptibility to probe orientation pose challenges. Future research should focus on standardizing imaging protocols and techniques to enhance its reproducibility and reliability. Additionally, longitudinal studies are needed to assess the predictive value of Doppler ultrasound in relation to long-term treatment outcomes and patient quality of life.

**CONCLUSIONS**

In conclusion, this study highlights the utility of Doppler ultrasound in evaluating the functional status of NVBs in prostate cancer patients undergoing RT. By providing insights into vascular dynamics and their correlation with sexual function outcomes, Doppler ultrasound has the potential to inform personalized treatment strategies and improve patient care. However, further research is warranted to address the study's limitations, validate the findings, and establish Doppler ultrasound as a valuable tool in the clinical management of prostate cancer.